\documentclass{article}

\def \be {\begin{equation}}
\def \ee {\end{equation}}
\def \bea {\begin{eqnarray}}
\def \eea {\end{eqnarray}}
\def \nn {\nonumber}

\def \a {\alpha}
\def \b {\beta}

\def \d {\delta}

\def \m {\mu}
\def \n {\nu}
\def \k {\kappa}

\def \s {\sigma}
\def \r {\rho}
\def \o {\omega}

\def \th {\theta}
\def \Th {\Theta}

\def \t {\tau}
\def \dag {\dagger}
\def \p {\partial}

\def\bd{\begin{document}}
\def\ed{\end{document}}
\def\nn{\nonumber}
\def\bea{\begin{eqnarray}}
\def\eea{\end{eqnarray}}
\let\bm=\bibitem
\let\la=\label

\def\N{{\cal N}}
\def\sst{\scriptscriptstyle}
\def\thetabar{\bar\theta}
\def\Tr{{\rm Tr}}
\def\one{\mbox{1 \kern-.59em {\rm l}}}

\def\a{\alpha}      \def\da{{\dot\alpha}}
\def\b{\beta}       \def\db{{\dot\beta}}
\def\c{\gamma}  \def\C{\Gamma}  \def\cdt{\dot\gamma}
\def\d{\delta}  \def\D{\Delta}  \def\ddt{\dot\delta}
\def\e{\epsilon}        \def\vare{\varepsilon}
\def\f{\phi}    \def\F{\Phi}    \def\vvf{\f}
\def\h{\eta}
\def\k{\kappa}
\def\l{\lambda} \def\L{\Lambda}
\def\m{\mu} \def\n{\nu}
\def\o{\omega}
\def\P{\Pi}
\def\r{\rho}
\def\s{\sigma}  \def\S{\Sigma}
\def\t{\tau}
\def\th{\theta} \def\Th{\Theta} \def\vth{\vartheta}
\def\X{\Xeta}
\def\z{\zeta}
\def\w{\wedge}
\def\u{\underline}
\def\hs{\hspace}

\def\cA{{\cal A}} \def\cB{{\cal B}} \def\cC{{\cal C}}
\def\cD{{\cal D}} \def\cE{{\cal E}} \def\cF{{\cal F}}
\def\cG{{\cal G}} \def\cH{{\cal H}} \def\cI{{\cal I}}
\def\cJ{{\cal J}} \def\cK{{\cal K}} \def\cL{{\cal L}}
\def\cM{{\cal M}} \def\cN{{\cal N}} \def\cO{{\cal O}}
\def\cP{{\cal P}} \def\cQ{{\cal Q}} \def\cR{{\cal R}}
\def\cS{{\cal S}} \def\cT{{\cal T}} \def\cU{{\cal U}}
\def\cV{{\cal V}} \def\cW{{\cal W}} \def\cX{{\cal X}}
\def\cY{{\cal Y}} \def\cZ{{\cal Z}}

\def\ua{\underline{\alpha}} \def\ubb{\underline{\beta}}
\def\ug{\underline{\gamma}}
\def\ub{\underline{\phantom{\alpha}}\!\!\!\beta}
\def\uc{\underline{\phantom{\alpha}}\!\!\!\gamma}
\def\um{\underline{\mu}} \def\un{\underline{\nu}}
\def\ud{\underline\delta}
\def\ue{\underline\epsilon}
\def\una{\underline a}\def\unA{\underline A}
\def\unb{\underline b}\def\unB{\underline B}
\def\unc{\underline c}\def\unC{\underline C}
\def\und{\underline d}\def\unD{\underline D}
\def\une{\underline e}\def\unE{\underline E}
\def\unf{\underline{\phantom{e}}\!\!\!\! f}\def\unF{\underline F}
\def\unm{\underline m}\def\unM{\underline M}
\def\unn{\underline n}\def\unN{\underline N}
\def\unp{\underline{\phantom{a}}\!\!\! p}\def\unP{\underline P}
\def\unq{\underline{\phantom{a}}\!\!\! q}
\def\unQ{\underline{\phantom{A}}\!\!\!\! Q}
\def\unH{\underline{H}}
\def\ul{\underline}

\def\As {{A \hspace{-6.4pt} \slash}\;}
\def\bs {{b \hspace{-6.4pt} \slash}\;}
\def\Ds {{D \hspace{-6.4pt} \slash}\;}
\def\ds {{\del \hspace{-6.4pt} \slash}\;}
\def\ss {{\s \hspace{-6.4pt} \slash}\;}
\def\ks {{ k \hspace{-6.4pt} \slash}\;}
\def\ps {{p \hspace{-6.4pt} \slash}\;}
\def\pas {{{p_1} \hspace{-6.4pt} \slash}\;}
\def\pbs {{{p_2} \hspace{-6.4pt} \slash}\;}

\def\Fh{\hat{F}}
\def\Vh{\hat{V}}
\def\Xh{\hat{X}}
\def\ah{\hat{a}}
\def\xh{\hat{x}}
\def\yh{\hat{y}}
\def\ph{\hat{p}}
\def\xih{\hat{\xi}}

\def\psit{\tilde{\psi}}
\def\Psit{\tilde{\Psi}}
\def\tht{\tilde{\th}}

\def\At{\tilde{A}}
\def\Qt{\tilde{Q}}
\def\Rt{\tilde{R}}
\def\Nt{\tilde{N}}

\def\at{\tilde{a}}
\def\st{\tilde{s}}
\def\ft{\tilde{f}}
\def\pt{\tilde{p}}
\def\qt{\tilde{q}}
\def\vt{\tilde{v}}
\def\nt{\tilde{n}}

\def\delb{\bar{\partial}}
\def\bz{\bar{z}}
\def\bD{\bar{D}}
\def\bB{\bar{B}}

\def\bk{{\bf k}}
\def\bl{{\bf l}}
\def\bp{{\bf p}}
\def\bq{{\bf q}}
\def\br{{\bf r}}
\def\bx{{\bf x}}
\def\by{{\bf y}}
\def\bR{{\bf R}}
\def\bV{{\bf V}}

\def\d{\delta}\def\D{\Delta}\def\ddt{\dot\delta}

\def\p{\partial} \def\del{\partial}
\def\xx{\times}
\def\uno{\mbox{1 \kern-.59em {\rm l}}}

\def\trp{^{\top}}
\def\inv{^{-1}}
\def\dag{{^{\dagger}}}

\def\pr{\prime}
\def\rar{\rightarrow}
\def\lar{\leftarrow}
\def\lrar{\leftrightarrow}

\def\cp{{\bf CP}^3}

\begin{document}
\title{Splitting of Folded Strings in $AdS_4\times CP^3$}
\author{Jun-Bao Wu$^{1, 2}$\footnote{Email: wujb@ihep.ac.cn}\\
\small{$^1$Institute of High Energy Physics,
and Theoretical Physics Center for Science Facilities,}\\
\small{ Chinese Academy of Sciences,
Beijing 100049, P.R. China}\\
\small{$^2$ Kavli Institute for Theoretical Physics China,
CAS, Beijing 100190, P.R. China}}
\date{\today}
\maketitle


\begin{abstract}
We study classically splitting of two kinds of folded string solutions in $AdS_4\times {\bf CP}^3$. Conserved charges of the
produced fragments are computed for each case. We find interesting patterns among these conserved charges.
\end{abstract}

\section{Introductions}
Quantization of string theory in $AdS_5\times S^5$ background is very important for the study of $AdS_5/CFT_4$ correspondence
\cite{Mal97, Gubser:1998bc, Witten:1998qj}. In spite of many efforts \cite{Berkovits:2000fe, Berkovits:2007rj, Bonelli:2008rv}, this is still very challenging even for free strings, due to the existence of the Ramond-Ramond fluxes. This difficulty led people to study Penrose limit of this background \cite{BMN} and semi-classical string solutions \cite{GKP} inside it. Largely inspired by these studies, people found
remarkable integrable structure in the planar limit \cite{Minahan:2002ve, Bena:2003wd} (for a collection of reviews, see \cite{Beisert:2010jr}).  With integrability at hand, people can compute many non-trivial quantities. Cusp anomalous dimension at arbitrary coupling \cite{Freyhult:2010kc} and scattering amplitudes at strong coupling \cite{Alday:2010kn} are two important examples of these quantities. Later this integrable structure was also found \cite{Minahan:2008hf, Bak:2008cp, Arutyunov:2008if, Stefanski:2008ik, arXiv:0811.1566, arXiv:1009.3498, arXiv:1101.3777} in $AdS_4/CFT_3$ correspondence \cite{Aharony:2008ug}.  The latter correspondence states that type IIA string theory in $AdS_4\times \cp$ is dual to certain three-dimensional ${\cal N}=6$ Chern-Simons-matter theory with gauge group $U(N)\times U(N)$ and Chern-Simons levels $(k, -k)$. For reviews on integrability in this correspondence, see \cite{Klose:2010ki, Lipstein:2011}.

Planar approximation in the gauge theory is dual to free approximation in the string theory side. Studying the interaction of strings in these nontrivial backgrounds is quite attracting and difficult. A single closed string can split into two because of interactions. One necessary condition for the splitting is that there are two points in the string which have the same positions and velocities. Folded strings satisfy this condition in a very simple manner. 
The splitting in flat spacetime can be studied at the full quantum level \cite{Iengo:2003ct}. However in backgrounds like $AdS_5\times S^5$, currently we can only study the splitting at the classical level  \cite{Peeters:2004pt, Murchikova:2011ea, Vicedo:2011vn}. The splitting of strings in $AdS_5\times S^5$ was studied in the gauge theory side in \cite{Peeters:2004pt, Casteill:2007td}. People hope that studies of the splitting will help us to improve our understanding about the interactions of strings in these non-trivial backgrounds and the behaviors of non-planar corrections in the gauge theory side \cite{Beisert:2002bb, Beisert:2003tq} ( for related review, please see \cite{Kristjansen:2010kg}) \footnote{Some discussions on non-planar sector via other approaches appeared in \cite{deMelloKoch:2009zm, Carlson:2011hy, Koch:2011jk, deMelloKoch:2011ci}.}.

Among other classical string solutions, two folded strings in $AdS_4\times \cp$ background were found in \cite{Chen:2008qq}. In this paper, we will study the splitting of these two strings. We compute their energy and the conserved charges from the isometry group $SU(4)$ of the $\cp$ part of the background geometry, both before \footnote{The energy and the Noether charges from the Cartan generators of $SU(4)$ before splitting have already been computed in \cite{Chen:2008qq}.} and after the splitting.  Some quite interesting patterns for these charges emerge for both folded strings. For the first folded string, we find that the
generators giving nonzero charges all belong to an $SU(3)$ subgroup of $SU(4)$. For the second folded string, we find that the conserved has the same form as the charges given in \cite{Peeters:2004pt} for splitting string in $AdS_5\times S^5$. The latter pattern can be thought as the extension of the similarity between the charges before the splitting already noticed in \cite{Chen:2008qq}. These similarities are quite interesting since there are many differences between $AdS_5/CFT_4$ and $AdS_4/CFT_3$ correspondences on both string theory side and gauge theory side. The study here on the concrete examples should be complementary to more abstract approach in \cite{Vicedo:2011vn}. Some aspects on non-planar corrections in the gauge theory side was discussed in \cite{Kristjansen:2008ib}.   

This paper is organized as follows: after discussing the Noether charges from the isometry group of $\cp$ in the next section, we study the splitting of two kinds of folded strings one by one in the following two sections. The last section is devoted to conclusions and discussions. We list the generators of the isometry group $SU(4)$ of $\cp$ in the fundamental representation in the appendix.

\section{Conserved charges}
In this paper, we focus on type IIA string theory in $AdS_4\times {\bf CP}^3$ background. The background metric is:
\be ds^2=R^2(\frac14 ds^2_{AdS_4}+ds^2_{{\bf CP}^3}),\ee
where $ds^2_{AdS_4}$ is metric of $AdS_4$ with unit radius
\be ds^2_{AdS_4}=- \cosh^2\rho dt^2+d\rho^2+\sinh^2\rho(d\theta^2+\sin^2\theta d\phi^2),\ee and the $ds^2_{{\bf} CP^3}$ is Fubini-Study metric of $\cp$ whose
details will be given below. The relation between $R$ and 't Hooft coupling $\lambda=N/k$ in the gauge theory side is:
\be R=2^{5/4}\pi^{1/2}\alpha^{\prime 1/2}\lambda^{1/4}, \label{lambda}\ee 
The dilaton field is a constant in this background:
\be e^{2\phi}=2^{5/2}\pi N^{1/2}k^{-5/2}.\ee There are also Ramond-Ramond background fields, but they do not play any roles in the studies here.

For the study of splitting classical string inside ${\bf CP}^3$, we would like to compute the conserved charges of these strings.
We start with the $\sigma$-model action:
\be S=\frac{1}{4\pi\alpha^\prime}\int \sqrt{-g}g^{\a\b}\p_\a X^\m \p_\b X^\n G_{\m\n}.\ee
The Killing vector $v^\mu$ of the target space gives a symmetry of the above action. And the corresponding
Noether current is:
\be j^\a=\frac1{2\pi\alpha^\prime} \sqrt{-g}g^{\a\b}\p_\b X^\n v^\m G_{\m\n}.\ee
We take the conformal gauge in which $g_{\a\b}$ is proportional to $\eta_{\a\b}=diag(-1, 1)$. Then we have
\be j^\a=\frac1{2\pi\alpha^\prime} \eta^{\a\b}\p_\b X^\n v^\m G_{\m\n}.\ee

${\bf CP}^3$ has isometry group $SU(4)$. The corresponding Killing vectors were given in \cite{Hoxha:2000jf} and more
explicitly in the coordinates we will use in \cite{Huang:2010qy}.
Let us begin with an unit $S^7$ \be\sum_{i=1}^4 |Z_i|^2=1,\ee inside ${\bf C}^4$. First define
\be\Omega_i=\sum_{I, J=1}^{4}(T_i)_I^{\,\,J}Z^IZ^\dagger_J \ee
with $T_i$ generators of $SU(4)$ in fundamental representation \footnote{They are listed in Appendix A.}.
Then the Killing vectors of ${\bf CP}^3$ are given by
\be v^\mu_i=J^{\mu\nu}\p_\n \Omega_i,\ee
where $J$ is the K\"ahler form of $\cp$.
From the above, we can get that
\be j^0_i=-\frac1{2\pi\alpha^\prime}\p_0 X_\mu J^{\mu\nu}\p_\nu \Omega_i.\ee

Now we choose a explicit system of coordinates for 
$\cp$.
We begin with the following parameterization of $S^7$:
 \bea
Z_1&=&\cos\xi\cos\frac{\theta_1}2\exp[i(y+\frac{\psi+\varphi_1}2)],\\
Z_2&=&\cos\xi\sin\frac{\theta_1}2\exp[i(y+\frac{\psi-\varphi_1}2)],\\
Z_3&=&\sin\xi\cos\frac{\theta_2}2\exp[i(y+\frac{-\psi+\varphi_2}2)],\\
Z_4&=&\sin\xi\sin\frac{\theta_2}2\exp[i(y+\frac{-\psi-\varphi_2}2)],\eea
where $0\le\xi<\frac\pi2, 0\le y<2\pi, -\pi\le\psi<\pi, 0\le\theta_i\le\pi,
0\le\varphi_i<2\pi$. 
Now the
induced metric on $S^7$ can be written as a $U(1)$ fiber over
$CP^3$: \be ds^2_{S^7}=ds^2_{CP^3}+(dy+A)^2. \ee Here $A$ is a one-form,
and the metric on ${\bf CP}^3$ is:
\bea ds^2_{{\bf CP}^3}&=&d\xi^2+\cos^2\xi\sin^2\xi
(d\psi+\frac12\cos\theta_1d\varphi_1-\frac12\cos\theta_2d\varphi_2)^2+\nn\\
&&\frac14\cos^2\xi(d\theta_1^2+\sin^2\theta_1
d\varphi_1^2)+\frac14\sin^2\xi(d\theta_2^2+\sin^2\theta_2 d\varphi_2^2). \eea
The K\"ahler form in this coordinate is \footnote{We choose the normalization of $J$ such that $J^\m_{\,\,\n} J^\n_{\,\,\rho}=-\delta^\m_\rho$.
}:
\bea J&=&-\frac12\sin\xi\cos\xi d\xi\wedge (2d\psi+\cos\theta_1d\varphi_1-\cos\theta_2d\varphi_2)-\frac14\cos^2\xi\sin\theta_1
d\theta_1\wedge d\varphi_1\nn\\
&& -\frac14 \sin^2\xi \sin\theta_2d\theta_2\wedge d\varphi_2.\eea

For the solutions consider in this paper, the $\tau$-dependent coordinates of ${\bf CP}^3$ are:
\be \psi=\omega_1\tau, \varphi_1=\omega_2\tau, \varphi_2=\omega_3\tau, \label{tau} \ee
then \bea j^0_i&=&-\frac{R^2}{2\pi\alpha^\prime}(\omega_1 J_{\psi}^{\,\,\,\,\nu}\p_\nu\Omega_i
+\omega_2 J_{\varphi_1}^{\,\,\,\,\nu}\p_\nu\Omega_i+\omega_3 J_{\varphi_2}^{\,\,\,\,\nu}\p_\nu\Omega_i)\\
&=&-\frac{\sqrt{\tilde\lambda}}{2\pi}(\omega_1 J_{\psi}^{\,\,\,\,\nu}\p_\nu\Omega_i
+\omega_2 J_{\varphi_1}^{\,\,\,\,\nu}\p_\nu\Omega_i+\omega_3 J_{\varphi_2}^{\,\,\,\,\nu}\p_\nu\Omega_i),\eea
where in the last line eq.~(\ref{lambda}) is used and we have defined that \be \tilde{\lambda}\equiv 32\pi^2\lambda. \ee

\section{Splitting of Folded String I}\label{folded1}

The first folded string solution we will study has the following configuration \cite{Chen:2008qq}:
\bea & & t=\kappa\tau, \rho=0, \theta_1=\theta_2=0,\\
& & \psi=\omega_1\tau, \varphi_1=\omega_2\tau, \varphi_2=\omega_3\tau,\eea
with $\xi=\xi(\sigma)$ being the only nontrivial function of $\sigma$.
When $\sigma$ increases from $0$ to $\pi$, $\xi$ increases from $-\xi_0$ to $\xi_0$ \footnote{Here when $\xi<0$, we change the coordinates $\xi, \varphi_2$ into $-\xi, \varphi_2+2\pi$.\label{footnote}}. While when $\sigma$ increases
from $\pi$ to $2\pi$, $\xi$ decreases from $\xi_0$ to $-\xi_0$.  The Virasoro constraints gives:
\be \frac{\kappa^2}{4}=\xi^{\prime
2}+\frac{\sin^22\xi}{4}\tilde{\omega}^2. \ee
where $\tilde{\omega}=\omega_1+(\omega_2-\omega_3)/2$ 
(we assume $\tilde{\omega}$ is positive in the remaining part
of this section). When $\xi=\xi_0$, we have $\xi^\prime=0$, then we get $\kappa^2=\sin^22\xi_0\tilde\omega^2$.
From this we have \be \xi^{\prime2}=\frac{\tilde \omega^2}4 (\sin^22\xi_0-\sin^22\xi).\label{xi}\ee
Then we have \be
2\pi=4\int_0^{\xi_0}\frac{2d\xi}{\tilde{\omega}\sqrt{\sin^22\xi_0-\sin^22\xi}},\ee
which gives \footnote{We use $E(q), K(q)$ to denote the elliptic integrals
of first kind and second kind, respectively. And we use $E(q, x)$ and $F(q, x)$ to denote the corresponding incomplete elliptic integral. Our
convention is that $E(q, \pi/2)=E(q), F(q, \pi/2)=K(q)$.}  \be \tilde{\omega}=\frac2{\pi}K(q) \ee with  $q=\sin^22\xi_0$.

Now we compute the conserved charges of this string before splitting. The energy is related
to translation along $t$ direction insider $AdS_4$ and is give by:
\be E
=\frac14\cosh^2\rho\sqrt{\tilde{\lambda}}\kappa.\ee
For this solution, the energy becomes:
\be E=\frac{\sqrt{\tilde{\lambda}}}{4}\kappa=\frac{\sqrt{\tilde{\lambda}}}{4}\tilde{\omega}
\sin2\xi_0=\frac{\sqrt{\tilde{\lambda}}}{2\pi}\sin2\xi_0K(q). \ee

The other Noether charges can be computed from the currents given in the previous section
\be Q_i=\int_0^{2\pi}j^0_i d\sigma.\ee We find that all
charges corresponding to non-Cartan generators vanish. This means that the solution is in highest weight representation of the isometry
group of the background \cite{Arutyunov:2003za}.

For Killing vectors corresponding to Cartan generators, we have \be v_3=-4\p_{\varphi_1}, v_8=\frac1{\sqrt{3}} (-4\p_\psi+4\p_{\varphi_2}),
v_{15}=-\frac1{\sqrt{6}} (4\p_\psi+8\p_{\varphi_2}),\ee
The corresponding Noether charges are:\bea Q_3&=&\frac{\sqrt{\tilde{\lambda}}}{\pi}(K(q)-E(q)),\\
Q_8&=&\sqrt{3}Q_3,\\ Q_{15}&=&0. \eea

Now assume that this folded string splits at $\sigma=a\in(0, \pi)$ and becomes two closed string. The first fragment
corresponds to $\sigma\in [0, a]\cup [2\pi-a, 2\pi]$ while the second fragment corresponds to $\sigma\in[a, 2\pi-a]$.
We denote $\xi(a)$ by $\xi_1$. From eq.~(\ref{xi}) The relation between $\xi_1$ and $a$ is:
\be \frac{\tilde\omega a}2=\int^{\xi_1}_{-\xi_0}\frac{d\xi}{\sqrt{\sin^22\xi_0-\sin^22\xi}},\ee
which gives:
\be a=\frac{1}{\tilde\omega}(K(q)+F(q, x_1)), \label{a}\ee
with \be x_1=\sin^{-1}\left(\frac{\sin2\xi_1}{\sin2\xi_0}\right)\ee

The energy of the first fragment is: \be E^I=\frac{a}\pi E=\frac{a\sqrt{\tilde{\lambda}}}{2\pi^2}\sin2\xi_0K(q).\label{ei}\ee
The other Noether charges for the first fragment is:
\be Q^I_{i}=2\int_0^a j^0_id\sigma. \ee
By some computations, we find that the Noether charges corresponding to the Cartan generators are:
\bea Q_3^I&=&\frac{\sqrt{\tilde{\lambda}}}{2\pi}(K(q)-E(q)+F(q, x_1)-E(q, x_1)),\label{q3i}\\
Q_8^I&=&\sqrt{3}Q_3^I,\\ Q_{15}^I&=&0. \eea
The non-vanishing Noether charges corresponding to the non-Cartan generators are:
\bea Q_4^I&=&\frac{\tilde\omega\sqrt{\tilde\lambda}}{2\pi}\cos\left(\frac12 (\omega_2-\omega_3+2\omega_1)\tau\right)\sqrt{q-\sin^22\xi_1} \\
 Q_5^I&=&\frac{\tilde\omega\sqrt{\tilde\lambda}}{2\pi}\sin\left(\frac12 (\omega_2-\omega_3+2\omega_1)\tau\right)\sqrt{q-\sin^22\xi_1}.  \eea
Without loss of generality, we can choose the time of splitting to be $\tau=0$, then using eq.~(\ref{tau}), we get:
\bea Q_4^I&=&\frac{\tilde\omega\sqrt{\tilde\lambda}}{2\pi} \sqrt{q-\sin^22\xi_1}, \label{q4i}\\
 Q_5^I&=&0.  \eea
For the charges of the second fragment we have $E^{II}=E-E^I$ and $Q^{II}_i=Q_i-Q^{I}_i, i=1, \cdots, 15$.
From eqs.~(\ref{a}, \ref{ei}, \ref{q3i}, \ref{q4i}) we can get relations among $E^I, Q^I_3$ and $Q^I_4$.
We notice that the non-vanishing charges belong to the set $Q^I_i, i=1, \cdots, 8$, with the corresponding
generators generate a $SU(3)$ subgroup of $SU(4)$.

\section{Splitting of folded string II}
The configuration of another folded string in \cite{Chen:2008qq} is:
\bea & & t=\kappa\tau, \xi=\pi/4, \theta_1(\sigma)=\pm\theta_2(\sigma),\\
& & \psi=\omega_1\tau, \varphi_1=\omega_2\tau, \varphi_2=\omega_3\tau,\eea
with $\omega_2=-\omega_3, \omega_1\neq 0$. Without losing generality, we assume that $\omega_1>0, \omega_2<0$, then the folded
string is around $\theta_1=0$.
When $\sigma$ increases from $0$ to $\pi$, $\theta_1$ increases from $-\theta_1^0$ to $\theta_1^{0}$ \footnote{Here, as in section~\ref{folded1}, when $\theta_1<0$, we change the coordinates $\theta_1, \theta_2, \psi, \varphi_1, \varphi_2$ into $-\theta_1, -\theta_2, \psi+\pi, \varphi_1-\pi, \varphi_2+\pi$. We thank the referee for suggestions on this footnote and footnote~\ref{footnote}.}. While when $\sigma$ increases
from $\pi$ to $2\pi$, $\theta_1$ decreases from $\theta_1^{0}$ to $-\theta_1^{0}$.
From the Virasoro constraint, we get:
\be \theta_1^{\prime2}=2\omega_1\omega_2(\cos\theta_1^0-\cos\theta_1), \ee
and \be \kappa^2=\omega_1^2+\omega_2^2+2\omega_1\omega_2\cos\theta_1^0.\ee
From \bea
2\pi=\int_0^{2\pi}d\sigma=\frac{2}{\sqrt{-2\omega_1\omega_2}}\int_0^{\theta_1(0)}\frac{d\theta_1}
{\sqrt{\cos\theta_1-\cos\theta_1^0}}, \eea
we get \be \sqrt{-\omega_1\omega_2}=\,\frac{2}{\pi}K(x),\ee
with $x=\sin^2\frac{\theta_1^0}{2}$.

The energy of this folded string is:
\be E=\frac14 \sqrt{\tilde\lambda}\kappa. \ee
As in last section,  the conserved charges corresponding to the non-Cartan generators vanish before the splitting of the string.
As for the ones for the Cartan generators we have:
\bea Q_3&=&-4Q_{\varphi_1},\\
Q_8&=&\frac1{\sqrt{3}}(-4Q_{\psi}+4Q_{\varphi_2}),\\
Q_{15}&=&-\frac1{\sqrt{6}}(4Q_{\psi}+8Q_{\varphi_2}).\eea
Now for this solution, we have \cite{Chen:2008qq}:
\bea
 Q_\psi
 &=&-\frac{\sqrt{\tilde{\lambda}}}{2\pi\sqrt{-\omega_1\omega_2}}\left((\omega_1-\omega_2)K(x)+2\omega_2E(x)
 \right),\\
Q_{\varphi_1}
&=&-\frac{\sqrt{\tilde{\lambda}}}{4\pi\sqrt{-\omega_1\omega_2}}\left((\omega_2-\omega_1)K(x)+2\omega_1E(x)\right),\\
Q_{\varphi_2}&=&-Q_{\varphi_1}. \eea
These charges satisfy the following relations:
\be\frac{\omega_1}{2}Q_{\psi}-\omega_2Q_{\varphi_1}=-\frac{\sqrt{\tilde{\lambda}}}{8}(\omega_1^2-\omega_2^2),
\ee
 \bea\label{relation}
 \left(\frac{E}{K(x)}\right)^2-\left(\frac{Q_\psi+2Q_{\varphi_1}}{E(x)}\right)^2&=&\frac{4\tilde\lambda x}{\pi^2},
 \\
 \left(\frac{Q_\psi-2Q_{\varphi_1}}{E(x)-K(x)}\right)^2-\left(\frac{Q_\psi+2Q_{\varphi_1}}{E(x)}\right)^2&=&\frac{4\tilde\lambda}{\pi^2}.
 \eea
As already pointed out in \cite{Chen:2008qq}, these relations are similar to the ones in \cite{Beisert:2003ea}. We will see that such kind of similarity appears after splitting as well.

 As in previous section, we assume that this folded string splits at $\sigma=a\in(0, \pi)$ and becomes two closed string. The first fragment
corresponds to $\sigma\in [0, a]\cup [2\pi-a, 2\pi]$ while the second fragment corresponds to $\sigma\in[a, 2\pi-a]$.
And we denote $\theta_1(a)$ by $\theta_1^1$. We can get the relation between $a$ and $\theta_1^1$ as:
\be a=\frac{K(x)+F(x, \tilde x)}{\sqrt{-\omega_1\omega_2}},\label{a2} \ee
with  \be\tilde x=\sin^{-1}\frac{\sin\frac{\theta_1^1}2}{\sin\frac{\theta_1^0}2}.\ee

The energy of the first fragment is: \be E^I=\frac{a}{\pi}E.\ee As for the charges corresponding to the Cartan generators, we have:
\bea Q_3^I&=&-4Q_{\varphi_1}^I,\\
Q_8^I&=&\frac1{\sqrt{3}}(-4Q_{\psi}^I+4Q_{\varphi_2}^I),\\
Q_{15}^I&=&-\frac1{\sqrt{6}}(4Q_{\psi}^I+8Q_{\varphi_2}^I),\eea with
\bea
 Q_\psi^I
 &=&-\frac{\sqrt{\tilde{\lambda}}}{4\pi\sqrt{-\omega_1\omega_2}}\left((\omega_1-\omega_2)(K(x)+F(x, \tilde x))\right.\nn\\
 &&\left. +2\omega_2(E(x)+E(x, \tilde x)
 \right),\\
Q_{\varphi_1}^I
&=&-\frac{\sqrt{\tilde{\lambda}}}{8\pi\sqrt{-\omega_1\omega_2}}\left((\omega_2-\omega_1)(K(x)+F(x, \tilde x))\right. \nn\\
&& \left.+2\omega_1(E(x)+E(x, \tilde x))\right),\\
Q_{\varphi_2}^I&=&-Q_{\varphi_1}^I. \eea

The nonzero charges for non-Cartan generators are:
\bea Q_1^I&=&-\frac{2\sqrt{\tilde\lambda}}{\pi}\frac{\omega_1}{\sqrt{-\omega_1\omega_2}}\cos(\omega_2\tau)\sqrt{\sin^2\frac{\theta^0_1}{2}-\sin^2\frac{\theta^1_1}{2}},\\
Q_2^I&=&-\frac{2\sqrt{\tilde\lambda}}{\pi}\frac{\omega_1}{\sqrt{-\omega_1\omega_2}}\sin(\omega_2\tau)\sqrt{\sin^2\frac{\theta^0_1}{2}-\sin^2\frac{\theta^1_1}{2}},\\
Q_6^I&=&\frac{2\sqrt{\tilde\lambda}}{\pi}\frac{\omega_2}{\sqrt{-\omega_1\omega_2}}\cos\left(\frac12 (\omega_2+\omega_3-2\omega_1)\tau\right)\sqrt{\sin^2\frac{\theta^0_1}{2}-\sin^2\frac{\theta^1_1}{2}},\\
Q_7^I&=&-\frac{2\sqrt{\tilde\lambda}}{\pi}\frac{\omega_2}{\sqrt{-\omega_1\omega_2}}\sin\left(\frac12 (\omega_2+\omega_3-2\omega_1)\tau\right)\sqrt{\sin^2\frac{\theta^0_1}{2}-\sin^2\frac{\theta^1_1}{2}},\\
Q_9^I&=&\pm\frac{2\sqrt{\tilde\lambda}}{\pi} \frac{\omega_2}{\sqrt{-\omega_1\omega_2}}\cos\left(\frac12 (\omega_2+\omega_3+2\omega_1)\tau\right)\sqrt{\sin^2\frac{\theta^0_1}{2}-\sin^2\frac{\theta^1_1}{2}},\\
Q_{10}^I&=&\pm \frac{2\sqrt{\tilde\lambda}}{\pi}\frac{\omega_2}{\sqrt{-\omega_1\omega_2}}\sin\left(\frac12 (\omega_2+\omega_3+2\omega_1)\tau\right)\sqrt{\sin^2\frac{\theta^0_1}{2}-\sin^2\frac{\theta^1_1}{2}},
\eea
the signs in the last two equations correlate with the sign in $\theta_1=\pm \theta_2$.

If we set the time of the splitting to be $\tau=0$, we get:
\bea Q_1^I&=&-\frac{2\sqrt{\tilde\lambda}}{\pi}\frac{\omega_1}{\sqrt{-\omega_1\omega_2}}\sqrt{\sin^2\frac{\theta^0_1}{2}-\sin^2\frac{\theta^1_1}{2}},\\
Q_2^I&=&0,\\
Q_6^I&=&\frac{2\sqrt{\tilde\lambda}}{\pi} \frac{\omega_2}{\sqrt{-\omega_1\omega_2}}\sqrt{\sin^2\frac{\theta^0_1}{2}-\sin^2\frac{\theta^1_1}{2}},\\
Q_7^I&=&0,\\
Q_9^I&=&\pm Q_6^I,\\
Q_{10}^I&=&0.
\eea

For the charges of the second fragment we still have $E^{II}=E-E^I$ and $Q^{II}_i=Q_i-Q^{I}_i, i=1, \cdots, 15$.

Now we notice that, interestingly, the results in eq.~(\ref{a2}) and the results for $E^I, Q_\psi^I, Q_{\varphi_1}^I, Q^I_1, Q^I_6$
have the same forms as the results for conversed charges in \cite{Peeters:2004pt} after suitably identifying some quantities here with some linear combinations of quantities there. Because of this relation, one can use the discussions in \cite{Peeters:2004pt} to get similar relations among these conserved charges before and after the splitting for the current case.

\section{Conclusions}

We study the splitting of one closed folded string in $AdS_4\times \cp$ into two for two cases at the classical level. We pay main attention to
the conserved charges after the splitting. In each cases, we find interesting pattern among these
conserved charges. For the first case, we find that the generators of the isometry group $SU(4)$ which lead to the non-vanishing Noether charges are in an $SU(3)$
subgroup of $SU(4)$. For the second case, the charges have the same forms as the ones in  \cite{Peeters:2004pt} for splitting of strings in $AdS_5\times S^5$.  It will be quite interesting to study reasons behind these patterns from both string theory and gauge theory sides.

The study here can be generalized into folded strings rotating in $AdS_4$ and $\cp$ at the same time following \cite{Murchikova:2011ea}.
It will be also interesting to study the higher conserved charges from the integrable structure of the strings,
using the methods in \cite{Eichenherr:1979mx, Scheler:1980yv, Chou:1980ym, Chou:1980zy}. We hope to address these issues in the future.

\section*{Acknowledgments}
JW would like to thank Bin~Chen and San-Min~Ke for discussions and encouragement.
The work of JW is partly supported by NSFC under Grant No. 11105154 by Youth Innovation Promotion Association, CAS.

\appendix
\section{The generators of $SU(4)$}
Now we list the generators of $SU(4)$ in fundamental representation used in \cite{Huang:2010qy} to compute the Killing vectors:
\bea
&&
T_1=\left(\begin{array}{cccc}
0&1&0&0\\
1&0&0&0\\
0&0&0&0\\
0&0&0&0
\end{array}\right),
T_2=\left(\begin{array}{cccc}
0&-i&0&0\\
i&0&0&0\\
0&0&0&0\\
0&0&0&0
\end{array}\right),
T_3=\left(\begin{array}{cccc}
1&0&0&0\\
0&-1&0&0\\
0&0&0&0\\
0&0&0&0
\end{array}\right),\nn\\
&&
T_4=\left(\begin{array}{cccc}
0&0&1&0\\
0&0&0&0\\
1&0&0&0\\
0&0&0&0
\end{array}\right),
T_5=\left(\begin{array}{cccc}
0&0&-i&0\\
0&0&0&0\\
i&0&0&0\\
0&0&0&0
\end{array}\right),
T_6=\left(\begin{array}{cccc}
0&0&0&0\\
0&0&1&0\\
0&1&0&0\\
0&0&0&0
\end{array}\right),\nn\\
&& T_7=\left(\begin{array}{cccc}
0&0&0&0\\
0&0&-i&0\\
0&i&0&0\\
0&0&0&0
\end{array}\right),
T_8=\frac1{\sqrt3}\left(\begin{array}{cccc}
1&0&0&0\\
0&1&0&0\\
0&0&-2&0\\
0&0&0&0
\end{array}\right),
T_9=\left(\begin{array}{cccc}
0&0&0&1\\
0&0&0&0\\
0&0&0&0\\
1&0&0&0
\end{array}\right),\nn\\
&&
T_{10}=\left(\begin{array}{cccc}
0&0&0&-i\\
0&0&0&0\\
0&0&0&0\\
i&0&0&0
\end{array}\right),
T_{11}=\left(\begin{array}{cccc}
0&0&0&0\\
0&0&0&1\\
0&0&0&0\\
0&1&0&0
\end{array}\right),
T_{12}=\left(\begin{array}{cccc}
0&0&0&0\\
0&0&0&-i\\
0&0&0&0\\
0&i&0&0
\end{array}\right),\nn\\
&&
T_{13}=\left(\begin{array}{cccc}
0&0&0&0\\
0&0&0&0\\
0&0&0&1\\
0&0&1&0
\end{array}\right),
T_{14}=\left(\begin{array}{cccc}
0&0&0&0\\
0&0&0&0\\
0&0&0&-i\\
0&0&i&0
\end{array}\right),
T_{15}=\frac1{\sqrt6}\left(\begin{array}{cccc}
1&0&0&0\\
0&1&0&0\\
0&0&1&0\\
0&0&0&-3
\end{array}\right).\nn
\eea


\begin{thebibliography}{99}

\bibitem{Mal97}
J.~M.~Maldacena,
``The large $N$ limit of superconformal field
theories and supergravity,''
Adv.\ Theor.\ Math.\ Phys.\ {\bf 2},
231 (1998), [hep-th/9711200].

\bibitem{Gubser:1998bc}
  S.~S.~Gubser, I.~R.~Klebanov and A.~M.~Polyakov,
 ``Gauge theory correlators from non-critical string theory,''
  Phys.\ Lett.\  B {\bf 428}, 105 (1998),
  [arXiv:hep-th/9802109].

\bibitem{Witten:1998qj}
  E.~Witten,
``Anti-de Sitter space and holography,''
  Adv.\ Theor.\ Math.\ Phys.\  {\bf 2}, 253 (1998),
  [arXiv:hep-th/9802150].

\bibitem{Berkovits:2000fe}
  N.~Berkovits,
  ``Super Poincare covariant quantization of the superstring,''
  JHEP {\bf 0004}, 018 (2000),
  [hep-th/0001035].

\bibitem{Berkovits:2007rj}
  N.~Berkovits, C.~Vafa,
``Towards a Worldsheet Derivation of the Maldacena Conjecture,''
  JHEP {\bf 0803}, 031 (2008),
  [arXiv:0711.1799 [hep-th]].

\bibitem{Bonelli:2008rv}
  G.~Bonelli, H.~Safaai,
  ``On gauge/string correspondence and mirror symmetry,''
  JHEP {\bf 0806}, 050 (2008),
  [arXiv:0804.2629 [hep-th]].

\bibitem{BMN}
  D.~E.~Berenstein, J.~M.~Maldacena and H.~S.~Nastase,
 ``Strings in flat space and pp waves from ${\cal N}=4$ super Yang Mills,''
  JHEP {\bf 0204}, 013 (2002),
 [arXiv:hep-th/0202021].


\bibitem{GKP}
  S.~S.~Gubser, I.~R.~Klebanov and A.~M.~Polyakov,
 ``A semi-classical limit of the gauge/string correspondence,''
  Nucl.\ Phys.\  B {\bf 636}, 99 (2002),
  [arXiv:hep-th/0204051].

\bibitem{Minahan:2002ve}
  J.~A.~Minahan, K.~Zarembo,
  ``The Bethe ansatz for ${\cal N}=4$ superYang-Mills,''
  JHEP {\bf 0303}, 013 (2003),
 [hep-th/0212208].

\bibitem{Bena:2003wd}
  I.~Bena, J.~Polchinski, R.~Roiban,
  ``Hidden symmetries of the $AdS_5 \times S^5$ superstring,''
  Phys.\ Rev.\  {\bf D69}, 046002 (2004),
  [hep-th/0305116].

\bibitem{Beisert:2010jr}
  N.~Beisert, C.~Ahn, L.~F.~Alday, Z.~Bajnok, J.~M.~Drummond, L.~Freyhult, N.~Gromov, R.~A.~Janik {\it et al.},
  ``Review of AdS/CFT Integrability: An Overview,''
   Lett.\ Math.\ Phys.\ {\bf 99}, 3 (2012), [arXiv:1012.3982 [hep-th]].

\bibitem{Freyhult:2010kc}
  L.~Freyhult,
  ``Review of AdS/CFT Integrability, Chapter III.4: Twist States and the cusp Anomalous Dimension,''
   Lett.\ Math.\ Phys.\  {\bf 99}, 255(2012), [arXiv:1012.3993 [hep-th]].

\bibitem{Alday:2010kn}
  L.~F.~Alday,
 ``Review of AdS/CFT Integrability, Chapter V.3: Scattering Amplitudes at Strong Coupling,''
 Lett.\ Math.\ Phys.\  {\bf 99}, 507(2012), [arXiv:1012.4003 [hep-th]].

\bibitem{Minahan:2008hf}
  J.~A.~Minahan, K.~Zarembo,
 ``The Bethe ansatz for superconformal Chern-Simons,''
  JHEP {\bf 0809}, 040 (2008),
 [arXiv:0806.3951 [hep-th]].

\bibitem{Bak:2008cp}
  D.~Bak, S.~-J.~Rey,
 ``Integrable Spin Chain in Superconformal Chern-Simons Theory,''
  JHEP {\bf 0810 } (2008)  053,
 [arXiv:0807.2063 [hep-th]].

\bibitem{Arutyunov:2008if}
  G.~Arutyunov, S.~Frolov,
  ``Superstrings on $AdS_4\times \cp$ as a Coset Sigma-model,''
  JHEP {\bf 0809 } (2008)  129,
  [arXiv:0806.4940 [hep-th]].

\bibitem{Stefanski:2008ik}
  B.~Stefanski, jr,
  ``Green-Schwarz action for Type IIA strings on $AdS_4\times \cp$,''
  Nucl.\ Phys.\  {\bf B808 } (2009)  80-87,
  [arXiv:0806.4948 [hep-th]].

\bibitem{arXiv:0811.1566}
  J.~Gomis, D.~Sorokin and L.~Wulff,
  ``The Complete $AdS_4\times \cp$ superspace for the type IIA superstring and D-branes,''
  JHEP\ {\bf 0903}, 015  (2009),
  [arXiv:0811.1566 [hep-th]].


\bibitem{arXiv:1009.3498}
  D.~Sorokin and L.~Wulff,
  ``Evidence for the classical integrability of the complete $AdS_4\times \cp$ superstring,''
  JHEP\ {\bf 1011}, 143  (2010),
  [arXiv:1009.3498 [hep-th]].


\bibitem{arXiv:1101.3777}
  D.~Sorokin and L.~Wulff,
  ``Peculiarities of String Theory on $AdS_4\times \cp$,''
  arXiv:1101.3777 [hep-th].

\bibitem{Aharony:2008ug}
  O.~Aharony, O.~Bergman, D.~L.~Jafferis, J.~Maldacena,
    ``${\cal N}=6$ superconformal Chern-Simons-matter theories, M2-branes and their gravity duals,''
  JHEP {\bf 0810}, 091 (2008),
  [arXiv:0806.1218 [hep-th]].


\bibitem{Klose:2010ki}
  T.~Klose,
   ``Review of AdS/CFT Integrability, Chapter IV.3: ${\cal N}=6$ Chern-Simons and Strings on $AdS_4 \times \cp$,''
   Lett.\ Math.\ Phys.\  {\bf 99}, 401 (2012), [arXiv:1012.3999 [hep-th]].

\bibitem{Lipstein:2011}
  A.~E.~Lipstein,
  ``Integrability of ${\cal N}=6$ Chern-Simons Theory,''
  [arXiv:1105.3231 [hep-th]].

\bibitem{Iengo:2003ct}
  R.~Iengo, J.~G.~Russo,
  ``Semiclassical decay of strings with maximum angular momentum,''
  JHEP {\bf 0303}, 030 (2003),
    [hep-th/0301109].

\bibitem{Peeters:2004pt}
  K.~Peeters, J.~Plefka, M.~Zamaklar,
   ``Splitting spinning strings in AdS/CFT,''
  JHEP {\bf 0411}, 054 (2004),
  [hep-th/0410275].

\bibitem{Murchikova:2011ea}
  E.~M.~Murchikova,
  ``Splitting of folded strings in $AdS_3$,''
  Phys.\ Rev.\  {\bf D84}, 026002 (2011),
  [arXiv:1104.4804 [hep-th]].

\bibitem{Vicedo:2011vn}
  B.~Vicedo,
  ``Splitting strings on integrable backgrounds,''
  [arXiv:1105.3868 [hep-th]].

\bibitem{Casteill:2007td}
  P.~-Y.~Casteill, R.~A.~Janik, A.~Jarosz, C.~Kristjansen,
  ``Quasilocality of joining/splitting strings from coherent states,''
  JHEP {\bf 0712}, 069 (2007),
  [arXiv:0710.4166 [hep-th]].

\bibitem{Beisert:2002bb}
  N.~Beisert, C.~Kristjansen, J.~Plefka, G.~W.~Semenoff and M.~Staudacher,
  ``BMN correlators and operator mixing in N = 4 super Yang-Mills theory,''
  Nucl.\ Phys.\  B {\bf 650}, 125 (2003),
  [arXiv:hep-th/0208178].

\bibitem{Beisert:2003tq}
  N.~Beisert, C.~Kristjansen and M.~Staudacher,
  ``The dilatation operator of N = 4 super Yang-Mills theory,''
  Nucl.\ Phys.\  B {\bf 664} (2003) 131,
  [arXiv:hep-th/0303060].

\bibitem{Kristjansen:2010kg}
  C.~Kristjansen,
  ``Review of AdS/CFT Integrability, Chapter IV.1: Aspects of Non-Planarity,''
 Lett.\ Math.\ Phys.\ {\bf 99}, 349 (2012), [arXiv:1012.3997 [hep-th]].

\bibitem{deMelloKoch:2009zm}
  R.~de Mello Koch, T.~K.~Dey, N.~Ives, M.~Stephanou,
  ``Hints of Integrability Beyond the Planar Limit: Nontrivial Backgrounds,''
  JHEP {\bf 1001}, 014 (2010),
  [arXiv:0911.0967 [hep-th]].

\bibitem{Carlson:2011hy}
  W.~Carlson, R.~d.~M.~Koch, H.~Lin,
  ``Nonplanar Integrability,''
  JHEP {\bf 1103}, 105 (2011),
  [arXiv:1101.5404 [hep-th]].

\bibitem{Koch:2011jk}
  R.~d.~M.~Koch, B.~A.~E.~Mohammed, S.~Smith,
  ``Nonplanar Integrability: Beyond the SU(2) Sector,''
  [arXiv:1106.2483 [hep-th]].

\bibitem{deMelloKoch:2011ci}
  R.~de Mello Koch, G.~Kemp and S.~Smith,
  ``From Large N Nonplanar Anomalous Dimensions to Open Spring Theory,''
  arXiv:1111.1058 [hep-th].

\bibitem{Chen:2008qq}
  B.~Chen, J.~-B.~Wu,
  ``Semi-classical strings in $AdS_4 \times \cp$,''
  JHEP {\bf 0809}, 096 (2008),
  [arXiv:0807.0802 [hep-th]].

\bibitem{Kristjansen:2008ib}
  C.~Kristjansen, M.~Orselli, K.~Zoubos,
  ``Non-planar ABJM Theory and Integrability,''
  JHEP {\bf 0903}, 037 (2009),
  [arXiv:0811.2150 [hep-th]].


\bibitem{Hoxha:2000jf}
  P.~Hoxha, R.~R.~Martinez-Acosta, C.~N.~Pope,
  ``Kaluza-Klein consistency, Killing vectors, and K\"ahler spaces,''
  Class.\ Quant.\ Grav.\  {\bf 17}, 4207-4240 (2000),
  [hep-th/0005172].

\bibitem{Huang:2010qy}
  Y.~-t.~Huang, A.~E.~Lipstein,
  ``Dual Superconformal Symmetry of N=6 Chern-Simons Theory,''
  JHEP {\bf 1011}, 076 (2010),
  [arXiv:1008.0041 [hep-th]].

\bibitem{Arutyunov:2003za}
  G.~Arutyunov, J.~Russo, A.~A.~Tseytlin,
  ``Spinning strings in $AdS_5\times S^5$: New integrable system relations,''
  Phys.\ Rev.\  {\bf D69}, 086009 (2004),
  [hep-th/0311004].

\bibitem{Beisert:2003ea}
  N.~Beisert, S.~Frolov, M.~Staudacher, A.~A.~Tseytlin,
  ``Precision spectroscopy of AdS / CFT,''
  JHEP {\bf 0310}, 037 (2003),
  [hep-th/0308117].

\bibitem{Eichenherr:1979mx}
  H.~Eichenherr,
   ``Infinitely Many Conserved Local Charges For The ${\bf CP}^{n-1}$ Models,''
    Phys.\ Lett.\  {\bf B90}, 121 (1980).

\bibitem{Scheler:1980yv}
  K.~Scheler,
   ``Local Coserved Currents For ${\bf CP}^N$ $\sigma$ Models,''
  Phys.\ Lett.\  {\bf B93}, 331 (1980).

\bibitem{Chou:1980ym}
  K.~-c.~Chou, X.~-c.~Song,
  ``Backlund Transformation, Local and Nonlocal Conservation Laws for Nonliear $\sigma$-Models on Symmetric Coset Spaces,''
  Sci.\ Sin.\  {\bf 25}, 716-722 (1982).

\bibitem{Chou:1980zy}
  K.~-c.~Chou, X.~-c.~Song,
  ``Local Conservation Laws for Various Nonliear $\sigma$-Models,''
  Sci.\ Sin.\  {\bf 25}, 825-833 (1982).

\end{thebibliography}
\end{document}